\documentclass[twocolumn,showpacs,prb,amsmath,amstex,amssymb,citeautoscript,longbibliography]{revtex4-1}
\pdfoutput=1
\usepackage{natbib}
\usepackage[english]{babel}
\usepackage{letltxmacro}
\usepackage{latexsym}
\LetLtxMacro{\ORIGselectlanguage}{\selectlanguage}
\makeatletter
\DeclareRobustCommand{\selectlanguage}[1]{%
  \@ifundefined{alias@\string#1}
    {\ORIGselectlanguage{#1}}
    {\begingroup\edef\x{\endgroup
       \noexpand\ORIGselectlanguage{\@nameuse{alias@#1}}}\x}%
}
\newcommand{\definelanguagealias}[2]{%
  \@namedef{alias@#1}{#2}%
}
\makeatother
\definelanguagealias{en}{english}
\definelanguagealias{English}{english}
\usepackage{graphicx}
\usepackage{amsmath}
\usepackage{amsfonts}
\usepackage{amssymb}
\usepackage{bm}
\usepackage{color}
\usepackage[percent]{overpic}
\usepackage{soul} 
\usepackage{amssymb}
\usepackage{wasysym}
\usepackage{dsfont}
\usepackage{float}

\usepackage{hyperref}
\hypersetup{
    bookmarks=false,         
    unicode=false,          
    pdftoolbar=false,        
    pdfmenubar=true,        
    pdffitwindow=false,     
    pdfstartview={FitH},    
    pdftitle={Thouless energy and multifractality across the many-body localization transition},    
    pdfauthor={Maksym Serbyn, Z. Papic, Dmitry A. Abanin},     
    pdfsubject={},   
    pdfcreator={},   
    pdfproducer={}, 
    pdfkeywords={many-body localization} {matrix elements} {disordered systems}, 
    pdfnewwindow=true,      
    colorlinks=true,       
    linkcolor=black,          
    citecolor=blue,        
    filecolor=magenta,      
    urlcolor=blue           
}

\newcommand{\be}{\begin{equation}}
\newcommand{\ee}{\end{equation}}
\newcommand{\bea}{\begin{eqnarray}}
\newcommand{\eea}{\end{eqnarray}}

\newcommand{\la}{\langle}
\newcommand{\ra}{\rangle}

\newcommand{\Eth}{{E_\text{Th}}}
\begin{document}
\title{Thouless energy and multifractality across the many-body localization transition}
\author{Maksym Serbyn$^1$, Z. Papi\'c$^2$, and  Dmitry A. Abanin$^3$}
\affiliation{$^1$ Department of Physics, University of California, Berkeley, California 94720, USA}
\affiliation{$^2$ School of Physics and Astronomy, University of Leeds, Leeds, LS2 9JT, United Kingdom}
\affiliation{$^3$ Department of Theoretical Physics, University of Geneva, 24 quai Ernest-Ansermet, 1211 Geneva, Switzerland}
\date{\today}
\begin{abstract}
Thermal and many-body localized phases are separated by a dynamical phase transition of a new kind. We analyze the distribution of off-diagonal matrix elements of local operators across the many-body localization transition (MBLT) in a disordered spin chain, and use it to characterize the breakdown of the eigenstate thermalization hypothesis and to extract the many-body Thouless energy. We find a wide critical region around the MBLT, where Thouless energy becomes smaller than the level spacing, while matrix elements show critical dependence on the energy difference. In the same region, matrix elements, viewed as amplitudes of a fictitious wave function, exhibit strong multifractality. Our findings show that the correlation length becomes larger than the accessible system sizes in a broad range of disorder strength values, and shed light on the critical behaviour of MBL systems.
\end{abstract}
\maketitle

The understanding of thermalization and its breakdown in isolated, interacting many-body systems is a fundamental and long-standing problem in quantum statistical mechanics. The recent strong theoretical interest in this problem is fuelled by the remarkable experimental advances, which led to the realization of synthetic, tunable systems of ultra-cold gases, where real-time quantum dynamics can be probed.~\cite{PolkovnikovRMP}

The fact that isolated ergodic systems, prepared in a non-equilibrium state, effectively reach thermal equilibrium is understood as a consequence of the Eigenstate Thermalization Hypothesis (ETH).~\cite{DeutschETH,SrednickiETH,RigolNature} The ETH states that observables, evaluated in individual eigenstates of an ergodic system, effectively take values given by the microcanonical ensemble, up to corrections that rapidly decay with the size of the system. In addition, to describe the approach to equilibrium, the ETH makes assumptions about the off-diagonal matrix elements of operators that describe physical observables, linking them with the macroscopic characteristics of the system.~\cite{Srednicki96,Srednicki99} 

Recently, many-body localization (MBL) has attracted much interest as a mechanism to break ergodicity and avoid thermalization.~\cite{Basko06,Mirlin05,OganesyanHuse,PalHuse} MBL phases exhibit the expected characteristics of a non-ergodic system: they violate the ETH, show area-law entanglement \cite{Bauer13,Serbyn13-1,Huse13}, and have the Poisson spectral statistics. These, as well as dynamical properties (including, e.g., a logarithmic spreading of entanglement~\cite{Znidaric08,Moore12,Serbyn13-2} and time dependence of local observables following a quench \cite{Serbyn14}), have been understood as a consequence of the emergence of Local Integrals of Motion (LIOMs) in the MBL phase.~\cite{Serbyn13-1,Huse13,Imbrie16,ScardicchioLIOM}

Generally, a quantum system is expected to exhibit a phase transition between the ergodic and localized phase as the disorder strength is increased. This is well known in the context of 
single-particle Anderson localization transition (ALT), which can be   characterized using the wave function of a system in real space. In the critical region, this wave function becomes multifractal. The Thouless energy -- the central concept in the scaling theory of Anderson localization, which sets the correlations between wave functions at different energies -- becomes comparable to the (single-particle) level spacing. 

Similarly, the many-body localization transition (MBLT) is driven by varying the strength of quenched disorder. While a complete theory of such a transition is lacking, phenomenological RG~\cite{AltmanRG14,Potter15X} and numerical~\cite{Demler14,Luitz-subdiff,Reichman15,Scard-15,Scard-16,Prelov-16,Gazit16}  studies have uncovered several fascinating properties in one-dimensional systems, notably the sub-diffusive transport on the ergodic side of the MBLT.  

Inspired by the successful description of the ALT using the Thouless energy and real-space wave functions, a question arises if similar concepts can be defined to describe the MBLT. One seemingly natural approach would be to view the many-body wave function as a single particle wave function in the Fock space, sites being non-entangled product states. The main difficulty with this approach is that the locality of the initial many-body Hamiltonian is lost, and physical observables become highly non-local in terms of the ``single particle'' wave function in the Fock space. 

In this work we present an alternative perspective. We show that matrix elements of the local operator $\hat O$ in the Hamiltonian eigenbasis can be viewed as an analogue of a wave function produced by the local excitation. First, by utilizing the relation between $O_{\alpha\beta}=\la \beta| \hat O|\alpha \ra$ (where $|\alpha\ra, |\beta\ra$ are the many-body eigenstates of the system) and dynamical properties of our system, we  extract the many-body Thouless energy. We discover that the Thouless energy becomes comparable to the many-body level spacing in the broad region preceding MBLT. We identify this region to be critical, similar to the case of ALT, and find that its width decreases as the system size is increased.

Second, treating matrix elements $O_{\alpha\beta}$ as amplitudes of a (fictitious) wave function obtained by acting on an eigenstate $|\alpha\rangle$ with a local operator $\hat O$, we study its statistics using the fractality toolbox developed in the context of Anderson localization.~\cite{Wegner1980,Soukoulis1984,Castellani1986,Evers2001,Rodriguez2010}  Recently similar approach was used in analytical studies of matrix elements within perturbation theory.~\cite{Montus-frac} In the same broad region surrounding the MBLT, where the Thouless energy becomes comparable to the level spacing, we find multifractal behavior of $O_{\alpha\beta}$, which suggests that MBLT corresponds to the ``freezing transition'' of the fractal spectrum. 

\begin{figure*}[t]
\begin{center}
\includegraphics[width=.66\columnwidth]{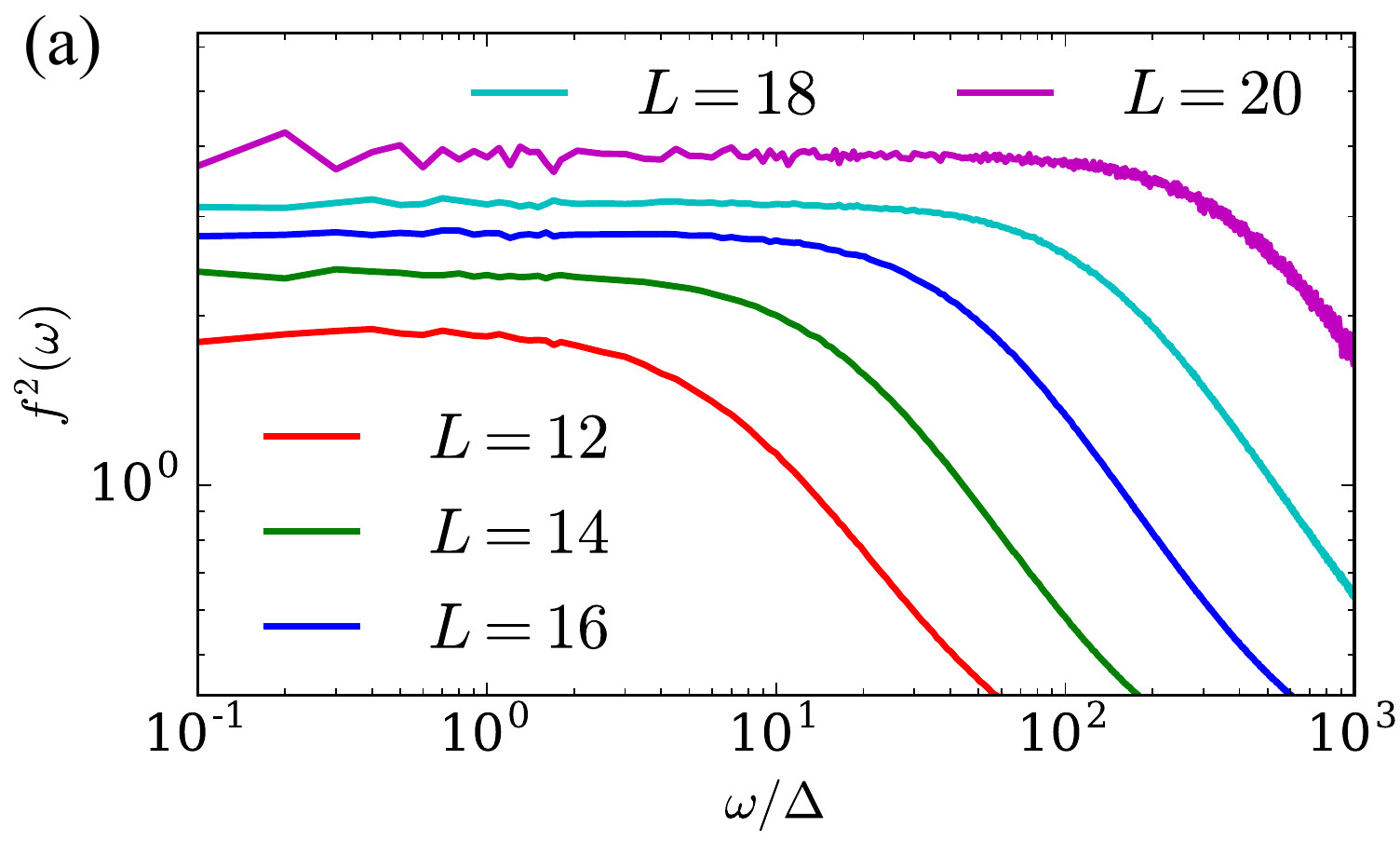}
\includegraphics[width=.66\columnwidth]{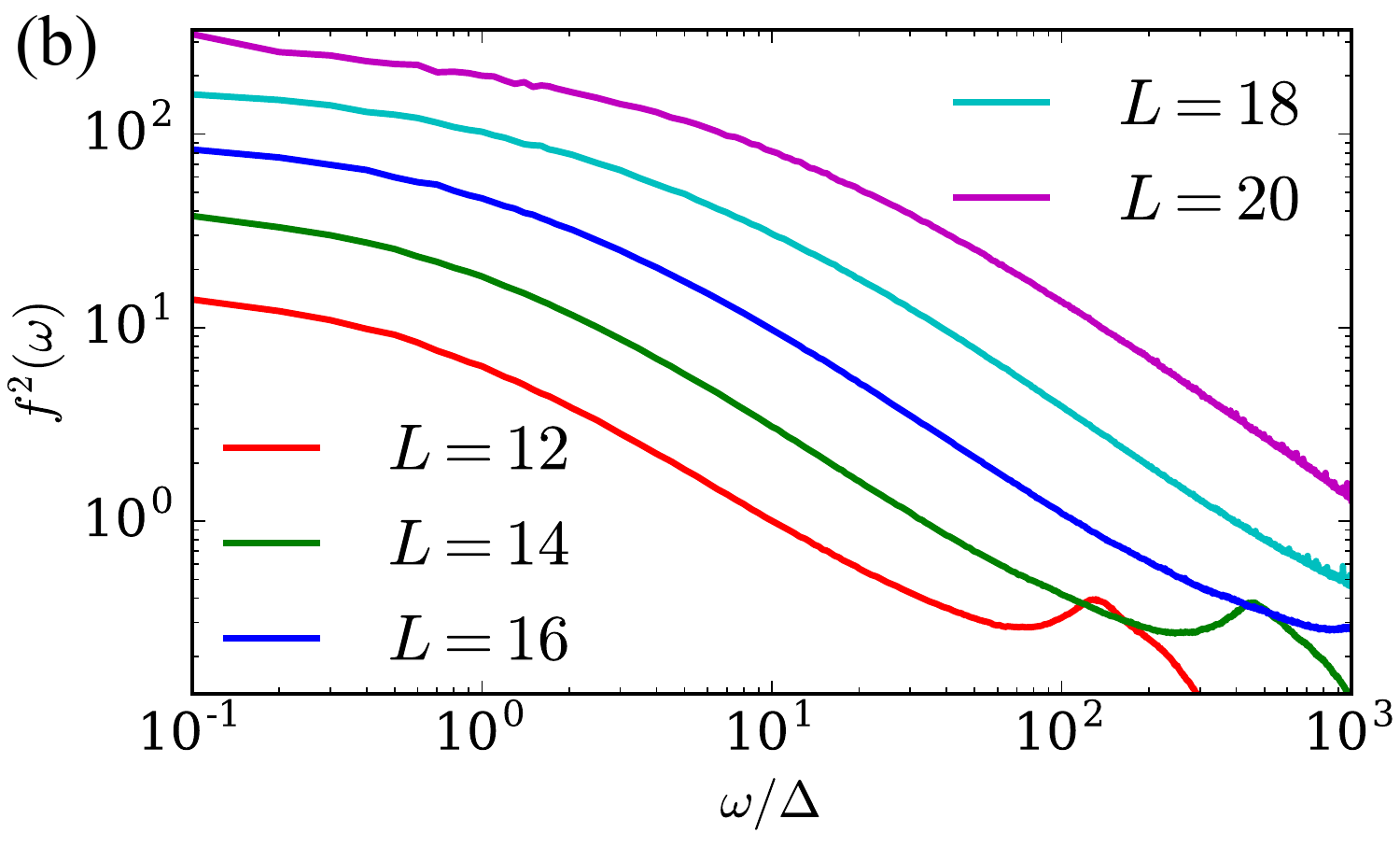}
\includegraphics[width=.66\columnwidth]{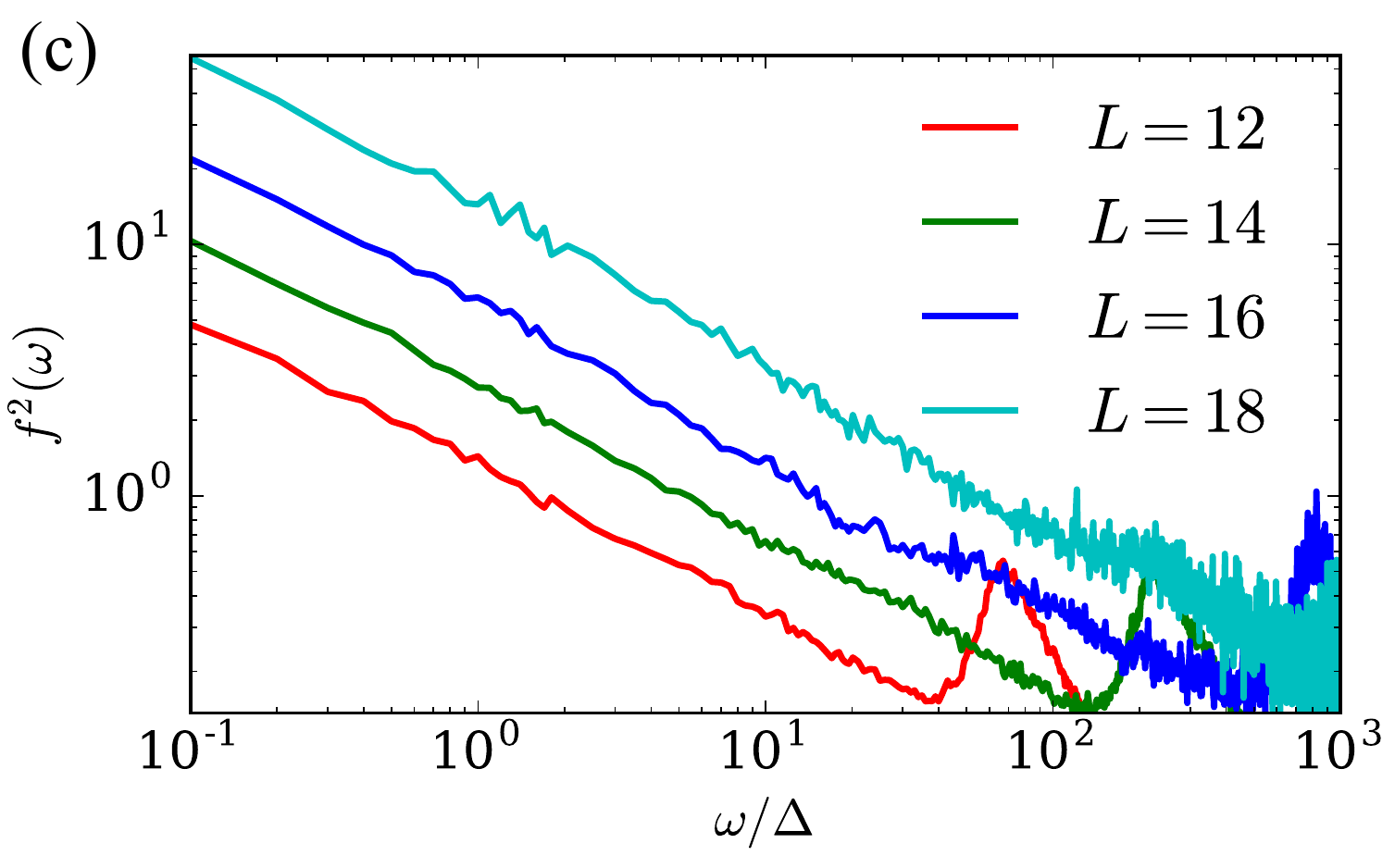}
\caption{ \label{Fig:f-local} (Color online) $f^2(\omega)$ is plotted as a function of $\omega/\Delta$, which probes the behavior at very low energies.  (a) For weak disorder $W=1$, the power-law decay of $f^2(\omega)$ is followed by the saturation at energies below $\Eth\gg \Delta$. Constant shift on the curves towards right with increasing system size is consistent with $\Eth$ scaling as a power-law in $L$. (b) For intermediate disorder $W=2$, the plateau fails to fully develop even for $L=20$ spins,  which signals that $\Eth\sim \Delta$. (c) In the MBL phase for $W=5$, $f^2(\omega)$ decays as a power-law for energies even below the many-body level spacing. }
\end{center}
\end{figure*}
 
{\it From matrix elements to dynamics.---} First, in order to study the dependence of matrix elements on the energy difference, we introduce a function 
\begin{equation}\label{eq:f-extr}
 f^2(\omega) 
 =
e^{S(E)}\langle  |O_{\alpha\beta}|^2 \delta(\omega-(E_\beta-E_\alpha))\rangle,
\end{equation}
where 
which is obtained by averaging  $|O_{\alpha\beta}|^2$, and $S(E)$ is the thermodynamic entropy at a given energy. 
This definition is inspired by the ETH~\cite{Srednicki99} ansatz for the matrix element in an ergodic system, $O_{\alpha\beta}={\mathcal O} (E)\delta_{\alpha\beta} +e^{-S(E)/2} f(E,\omega) R_{\alpha\beta}$,
where the first term corresponds to the expectation value of $\hat O$, which is a smooth function of the energy $E$.  The second term, describing off-diagonal matrix elements, is a product of random numbers $R_{\alpha\beta}$ with zero mean and unit variance, and a smooth function $f(\omega)$ that depends on the energy difference $\omega=E_\alpha-E_\beta$ and average energy, $E=(E_\alpha+E_\beta)/2$.  This ansatz has been tested numerically in several models.~\cite{Rigol-FDT,Prelov13,Haque15,Polkovnikov-rev} Assuming that $\delta$-function in Eq.~(\ref{eq:f-extr}) is broadened on a scale much larger than the many-body level spacing, and that $R_{\alpha\beta}$ averages out, we see that Eq.~(\ref{eq:f-extr}) reduces to the ETH ansatz. 

Function $f^2(\omega)$ encodes two important physical characteristics of our system. In particular, neglecting the variation of $S(E)$ near $E=0$, one can show  that $f^2(\omega)$ coincides with the average spectral function for states with energy $E_\alpha$ close to zero, which can be measured, e.g., in a tunnelling or absorption experiment. Furthermore, the Fourier transform of $f^2(\omega)$ determines the time dependence of the connected correlation function,~\cite{Polkovnikov-rev}  
\be\label{eq:correlator}
F_\alpha(t)=\la \alpha| \hat O(t) \hat O(0) |\alpha\ra_c\approx \int_{-\infty}^\infty d\omega\, e^{-i\omega t} f^2(\omega),
\ee
provided the $f(\omega)$ is a smooth function, and its fluctuations  average out.
Function $F_\alpha(t)$ enters the generalization of the Fluctuation-Dissipation Theorem,~\cite{Rigol-FDT,Polkovnikov-rev} along with the Kubo linear response. Finally, for the case when $\hat O$ corresponds to the density of a conserved quantity, $F(t)$ gives the return probability and can be related to conductivity via the Einstein relation.~\cite{Demler14} 

The correlation function~(\ref{eq:correlator}) can be accessed directly in numerical studies for short times in relatively large systems.~\cite{Luitz-subdiff} In order to characterize the long-time dynamics, below we study the function $f^2(\omega)$ at small frequencies. In particular, we will use it to directly extract the {\it many-body Thouless energy} which sets the time scale at which the dynamics saturate in a finite size system.~\cite{Polkovnikov-rev} We also note that the  function $f^2(\omega)$ was studied in a different context in prior work: transition in the spectral function was suggested to exist in the MBL phase,~\cite{Rahul14} it was related to the level statistics~\cite{Serbyn16}; recently, its fluctuations at very small $\omega$ were connected to the dynamical exponent.~\cite{Luitz-fluc-16}

{\it Model and methods.---}We focus on the 1D random-field spin-$1/2$ Heisenberg model,~\cite{PalHuse} 
\be\label{eq:Heisenberg}
H= \sum_i \mathbf{S}_i \mathbf{S}_{i+1} + h_i S_i^z,
\ee
where $S_i^\alpha = \sigma_i^\alpha/2$ is the spin operator on site $i$, expressed in terms of Pauli matrices $\sigma^\alpha$, and the magnitude of the random field is uniformly distributed $h_i\in [-W;W]$.
The only control parameter is the disorder strength,~$W$: the model~(\ref{eq:Heisenberg}) was demonstrated to be in the fully MBL phase at $W>W_c$.~\cite{PalHuse,Alet14,Serbyn15}  Eigenstates in the center of the many-body band become delocalized, and the mobility edge appears at values $W<W_c$. Previous studies determined the location of the transition, $W_c\approx 3.75$, and found that the transport is {\it sub-diffusive} on the delocalized side of the transition, $W<W_c$;~\cite{Demler14,Luitz-subdiff,Reichman15,Scard-15,Scard-16,Prelov-16,Gazit16} with some studies suggesting subdiffusive behavior down to very small disorder.~\cite{Reichman15,Gazit16} We will mostly work with the operator $\hat O=\sigma_i^z$; in the Appendix we provide similar data for other operators. 

In order to extract properties of the matrix elements, we use exact diagonalization (ED) for spin chains of length $L=10,\ldots, 16$ and shift-invert (SI) algorithm from  PETSc/SLEPc package~\cite{SLEP} with MUMPS eigensolver for $L=18,\ 20$  spins. We consider states in the total $S^z=0$ sector near zero energy and use from $10^4$ to $100$ disorder realization for $L=10,\ldots, 20$ spins. With SI algorithm we obtain $10^3$ eigenvectors closest to the target energy $E=0$. 

{\it Spectral function.---} We start by considering the function $f^2(\omega)$ defined in Eq.~(\ref{eq:f-extr}), with averaging performed both over eigenstates $\alpha, \beta$ and disorder realizations. 
In the Appendix, we show that $f^2(\omega)$  weakly depends on the system size unless $\omega\ll1$ (i.e., $\omega$ is much smaller than the microscopic energy scale which is one in our units). Moreover, due to locality of the operator $\hat O$, $f^2(\omega)$ decays exponentially for $\omega\geq 1$.~\cite{Abanin15}

In order to access the properties of $f^2(\omega)$ at low energies, which correspond to the long-time dynamics, we plot it as a function of $\omega/\Delta$, where $\Delta \approx W \sqrt{L}/{\cal D}$ is the many-body level spacing (${\cal D}(L) = {L\choose L/2}$ is the Hilbert space dimension). Fig.~\ref{Fig:f-local} shows the behavior of $f^2(\omega)$ for different system sizes for increasing disorder strength. Deep in the ergodic phase, Fig.~\ref{Fig:f-local}(a), $f^2(\omega)$ behaves as
\begin{equation}\label{Eq:phi-Eth}
  f^2(\omega) \propto \frac{1}{\omega^\phi},
  \quad
  \text{for}\quad
  \Eth\ll \omega \ll 1,
\end{equation}
saturating to a constant for energies $\omega\lesssim \Eth$. 
This is fully consistent with the expectation that the correlation function~(\ref{eq:correlator}) relaxes as a power law even in the ergodic phase because $\hat O=\sigma^z_i$ is the local density of a conserved quantity (the $z$ spin projection). Then, from the decay of $F(t)\propto 1/t^\gamma$, corresponding to diffusion for $\gamma=1/2$ and sub-diffusion for $\gamma<1/2$, and relation~(\ref{eq:correlator}), we expect $f^2(\omega)\propto 1/\omega^\phi$ with  $\phi=1-\gamma$. Furthermore, the (sub)diffusive dynamics are saturated at a time $t\sim t_* = L^{1/\gamma}$. Then, we expect the saturation of $f^2(\omega)$ for energies $\omega<\Eth$ with  $\Eth = 1/t_* \propto L^{-1/\gamma}$, as is indeed the case in Fig.~\ref{Fig:f-local}(a). The featureless form of $f^2(\omega)$ for $\omega<\Eth$ is natural since for times longer than $t_*$, a local excitation explores the full system size, and dimensionality is effectively lost -- the system is described by the random matrix theory. 

Notably, $\Eth$ rapidly decreases as we increase the disorder strength. In particular, already for $W=2$, still far from the MBLT at $W_c$, Fig.~\ref{Fig:f-local}(b) illustrates the absence of a fully developed plateau for the largest studied system sizes.  The slight upward curvature of $f^2(\omega)$ for  $\omega<\Delta$ persists through the MBLT up to disorder $W\lesssim 4$. At even stronger disorders, e.g., $W=5$ in Fig.~\ref{Fig:f-local}(c), the part of $f^2(\omega)$ with upward curvature entirely disappears, and $f^2(\omega)$ retains a power-law shape down to the energies below level spacing. Below we explore the variation of   $\Eth$ and power $\phi$  across the  MBLT.

\begin{figure}[b]
\begin{center}
\includegraphics[width=0.95\columnwidth]{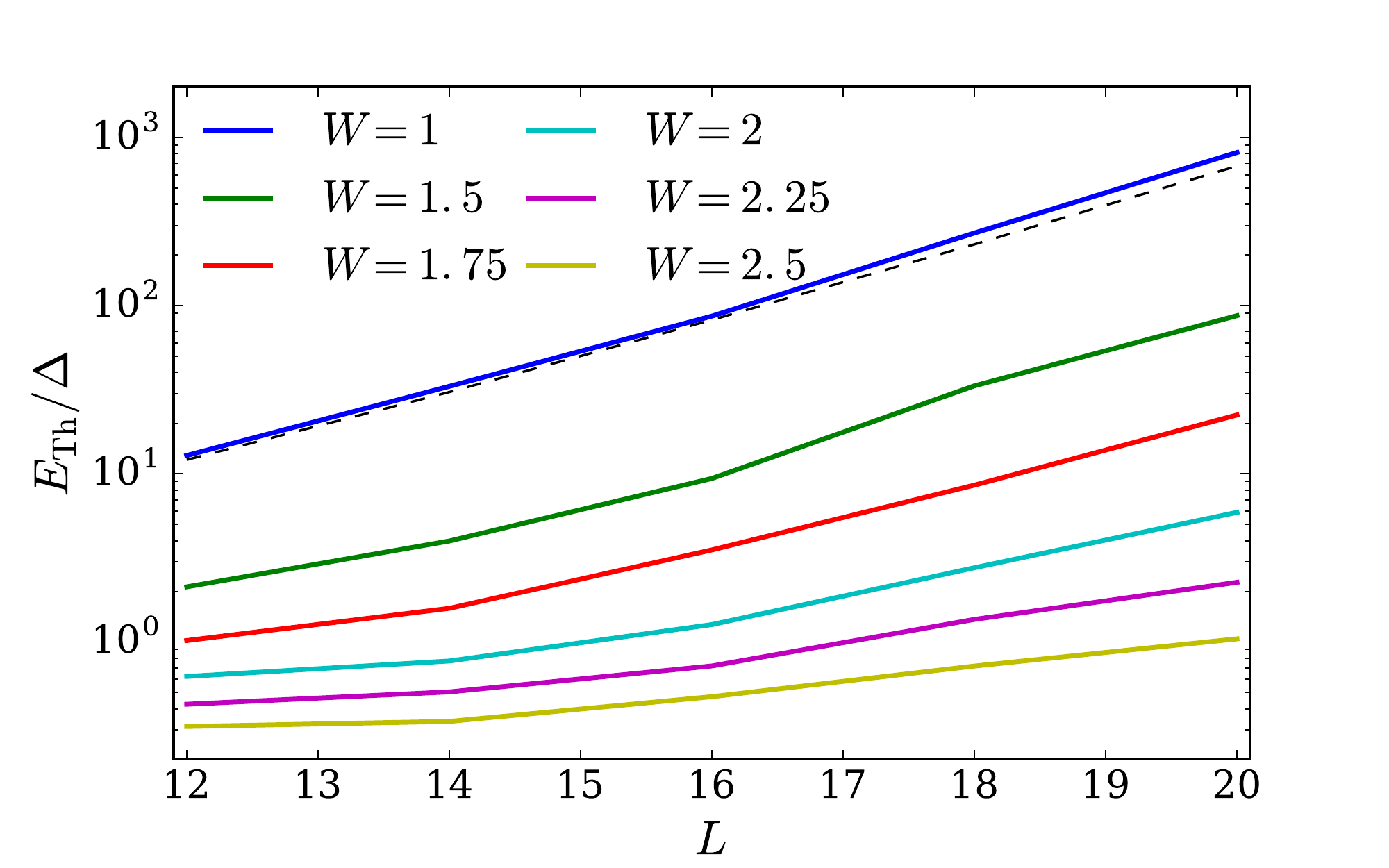}\\
\caption{ \label{Fig:Eth}
(Color online) Growth of  $\Eth/\Delta$ with $L$ slows down as the value of d the isorder is increased towards the MBLT. Already for $W=2.5$ we have $\Eth<\Delta$ even for $L=20$. For larger disorders $\Eth$ is too small to be reliably determined.}
\end{center}
\end{figure}

{\it Thouless energy.---}In order to systematically study the behavior of $\Eth$, we fit the corresponding curves for $f^2(\omega)$ with the function 
\begin{equation}\label{Eq:fit}
f^2(\omega)
=
\frac{f^2(0)}{1+(\omega/\Eth)^\phi}.
\end{equation}
Fig.~\ref{Fig:Eth} shows the Thouless energy extracted from such fits.  Dashed line in Fig.~\ref{Fig:Eth} corresponds to the diffusive ($\gamma=1/2$) scaling  of $\Eth\propto 1/L^2$,  and is roughly consistent with the data for weak disorder $W=1$. For disorder $W\geq W_*$, where $W_*\approx 2$, the growth of $\Eth/\Delta$ becomes increasingly slower. From Fig.~\ref{Fig:Eth} it is evident $\Eth$ remains below the level spacing $\Delta$ for $W>W_*$ and all available system sizes.  While one cannot rule out the power-law behavior $\Eth\propto L^{-1/\gamma}$ with very small $\gamma$,  it is more natural to  interpret the data as   \emph{exponential} dependence of $\Eth$ with $L$, $\Eth \propto e^{-\kappa L}$ with $\kappa<\ln 2$.

We interpret the small value $\Eth\leq \Delta$, and  exponential scaling of $\Eth$ as evidence for our system entering the critical region near the MBLT. Indeed, recent phenomenological RG studies~\cite{AltmanRG14,Potter15X} suggested a scaling $\log \tau \sim L$ in the critical region near the MBLT; logarithmic growth of particle number fluctuations for $W<W_c$ was also demonstrated numerically.~\cite{Serbyn15}  If there exists a correlation length $\xi(W)$ that depends on disorder and diverges at the MBLT for $W=W_c$, then even at disorder $W<W_c$ small systems of length $L\leq\xi(W)$ may qualitatively behave as if they were at the MBLT transition, so one has to study systems of size $L>\xi(W)$ to see the delocalized behavior. In this scenario, from Fig.~\ref{Fig:Eth} it follows that correlation length $\xi(W_*)\geq 20$ becomes larger than our largest system size at disorder value $W_*\approx 2$.~\footnote{Refs.~\onlinecite{AltmanRG14,Potter15X} suggested scaling $\gamma\sim1/\xi(W)$, which implies $\Eth\sim e^{-\alpha\xi(W) \log L}$. Hence, $\Eth$ becomes much smaller than level spacing $\sim e^{- L \ln 2}$ when $\xi(W)\gg L$, which matches the condition for being in the critical fan regime.} Below we present additional evidence for critical behavior from power $\phi$ governing the decay of $f^2(\omega)$.

\begin{figure}[t]
\begin{center}
\includegraphics[width=0.95\columnwidth]{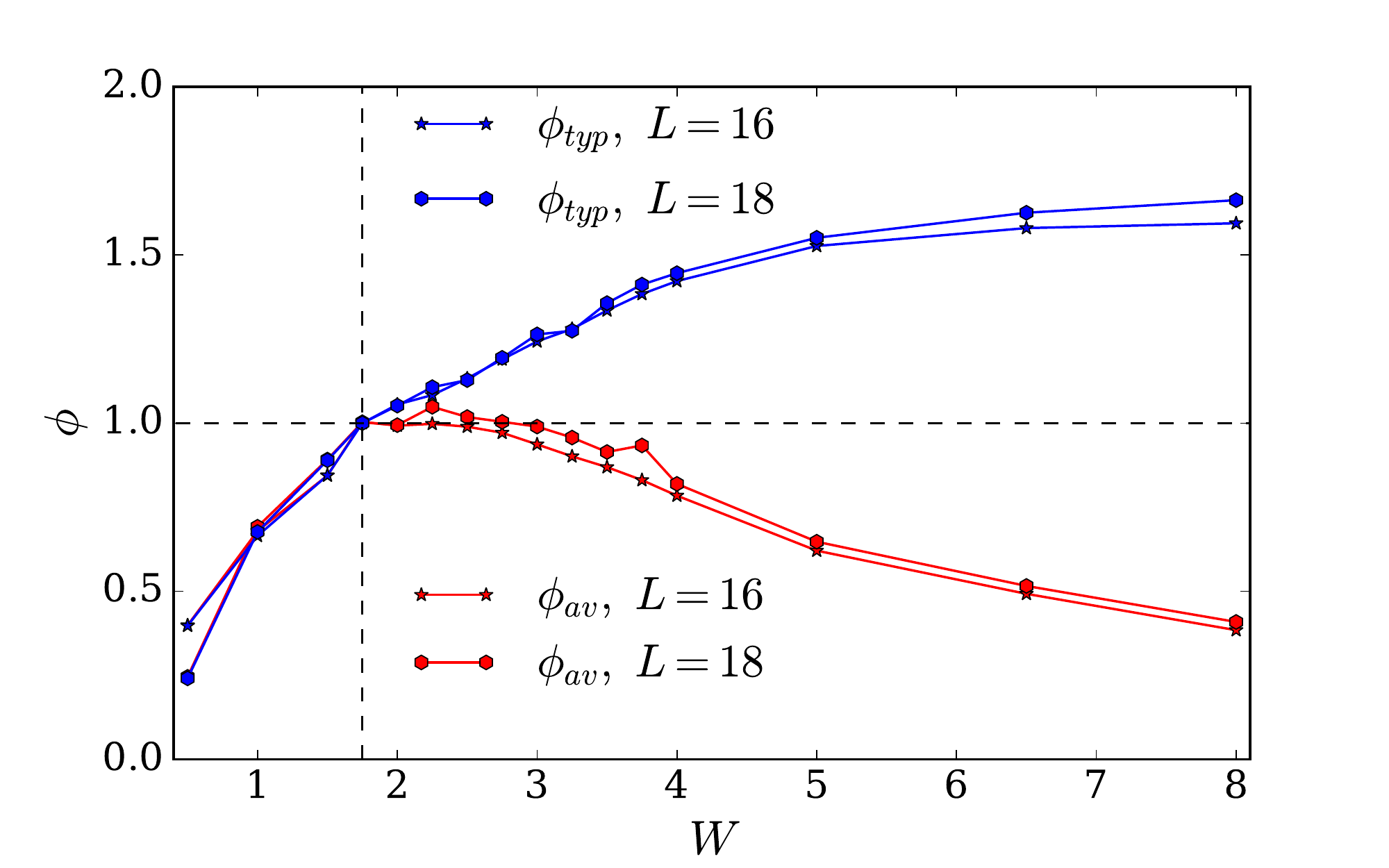}\\
\caption{ \label{Fig:phi} (Color online)
Powers $\phi_{av}$ and $\phi_{typ}$ controlling the decay of $f^2(\omega)$ and $[f^2(\omega)]_\text{typ}$ saturate to one near $W_*\approx 2$. For larger values of the disorder, $\phi_{av}$ decreases again, while the power $\phi_{typ}$ controlling the typical behavior grows above one.}
\end{center}
\end{figure}

{\it Power $\phi$ across the MBLT.---} The power $\phi$ extracted from fitting $f^2(\omega)$ to Eq.~(\ref{Eq:fit}) for disorder $W\geq 1.75$  is shown in Fig.~\ref{Fig:phi} (red lines). For smaller values of $W$,  Eq.~(\ref{Eq:fit}) gave satisfactory fits at small $\omega$ but failed to capture the abrupt onset of the power-law decay. Thus for  $W\leq 1.5$ we extracted the power $\phi_{av}$ from fitting the central part of the ``shoulder'' [see Fig.~\ref{Fig:f-local}(a)] to the power law. Blue curves in Fig.~\ref{Fig:phi} refer to $\phi_{typ}$ defined as the power-law governing the decay of log-averaged $[f^2(\omega)]_{typ}=\exp(\langle \ln f^2(\omega) \rangle)$, where the brackets denote averaging over disorder and eigenstates. 

From Fig.~\ref{Fig:phi} we observe that for disorders $W<W_*$ both powers coincide, and are smaller than one, $\phi_{typ}\approx \phi<1$. This is consistent with the subdiffusion observed numerically in the studies of different correlation functions in the time domain,~\cite{Demler14,Luitz-subdiff,Reichman15,Scard-15,Scard-16} also Refs.~\onlinecite{Gopa-15,Prelov-16} directly addressed conductivity in the frequency domain. For $W\geq W_*$, the power $\phi$ saturates to one for some range of the disorder values. This corresponds to  $f^2(\omega)$ decaying as $1/\omega$, and suggests a logarithmic decay of correlation functions in time. Such behavior again is consistent with the system entering the critical region for the accessible system sizes. 

Note, that for disorder $W>W_*$ power $\phi$ and $\phi_{typ}$ extracted from average and log-averaged spectral functions do not agree anymore. This signals that the function $f^2(\omega)$ is no longer smooth, and is consistent with the onset of strong multifractality in the critical fan, which we address below. The dominant contribution to $f^2(\omega)$ now comes from rare resonances that give matrix elements of order one.~\cite{Gopa-15} In contrast, the log-averaged spectral function is dominated by most probable matrix elements, and $[f^2(\omega)]_\text{typ}$   exponentially decreases with $L$, consistent with the criterion for the MBLT.~\cite{Serbyn15} 
\begin{figure}[t]
\begin{center}
\includegraphics[width=.99\columnwidth]{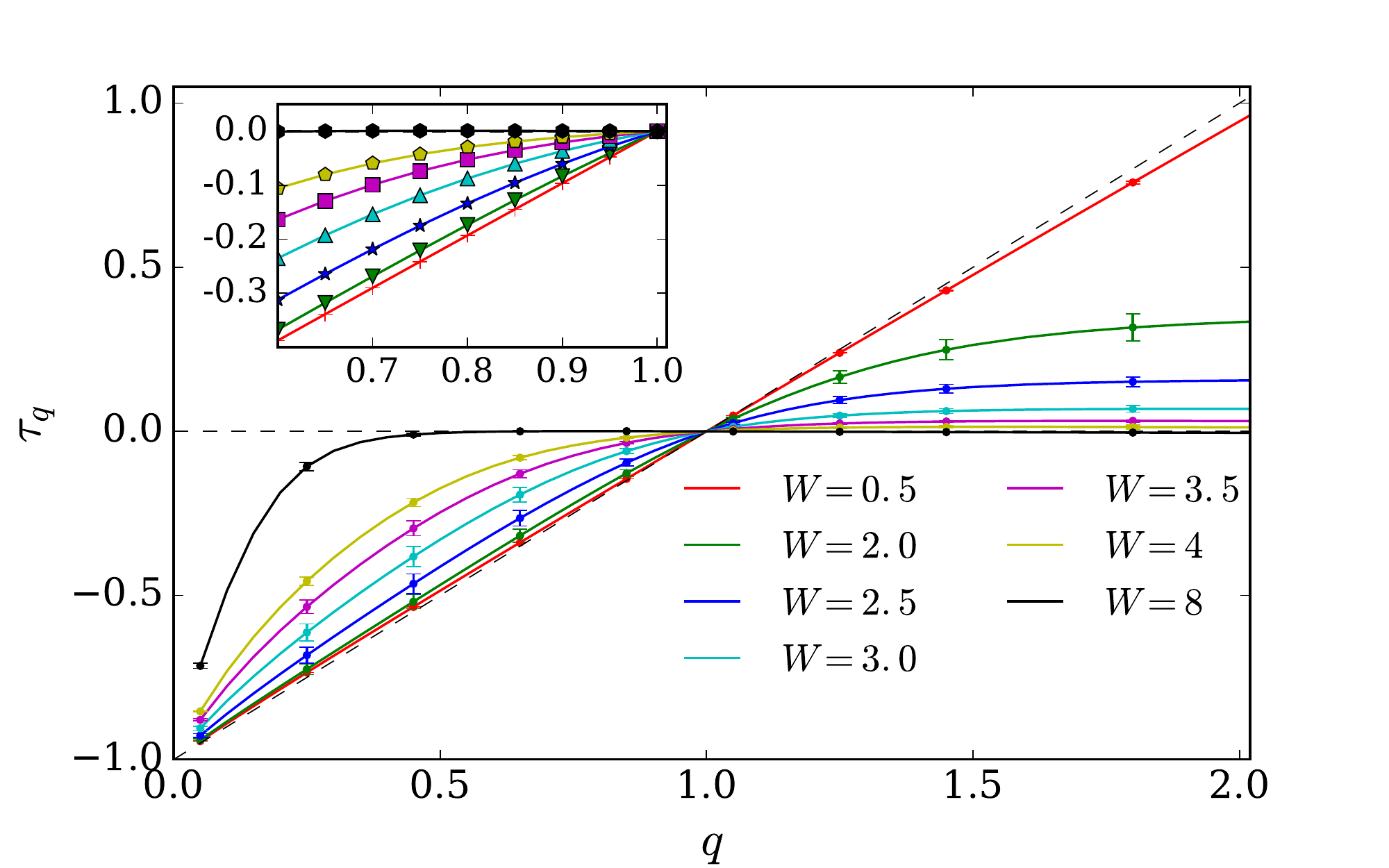}\\
\caption{ \label{Fig:IPR} (Color online) The spectrum of exponents $\tau_q$ governing the scaling of participation ratios $P_q$. The spectrum evolves from being very close to that of a metal (red line, $W=0.5$) to the ``frozen'' fractal spectrum deep in the MBL phase ($W=8$, black line). Inset shows $\tau_q$ near $q\leq 1$.}
\end{center}
\end{figure}

{\it Multifractal analysis of the matrix elements.---} Finally, we consider an interpretation of the matrix elements as the wave function amplitudes after applying a local perturbation. We locally perturb our system  with an operator $\hat O$ starting from an eigenstate $|\alpha\ra$. Then, the probability to find the system in the eigenstate $|\beta\ra$  is given by  $|O_{\alpha\beta}|^2=|\psi_\alpha(\beta)|^2$, which can be interpreted as the squared wave function amplitude. When the operator $\hat O$ squares to the identity, as is the case for $\hat O = \sigma^z_i$, the wave function is normalized, $\sum_\beta |\psi_\alpha(\beta)|^2 = 1$.

 Although the fractality was recently argued to be a generic property of many-body ground states,~\cite{Alet-frac} fractal properties strongly depend on the choice of the basis. In particular, previous works~\cite{Luca13,Torres-15} studied the decomposition of excited many-body eigenstates in the standard basis of product states. Here, in contrast, we work in the eigenbasis of the unperturbed Hamiltonian, and relate fractal dimensions in this basis to the onset of criticality and localization. 

It is natural to expect that the statistics of this wave function would provide useful information about the localization/ergodicity properties of the system. In order to understand the fractal properties of matrix elements, we study participation ratios with general index $q$, $P_q$, of the wave function $\psi_\alpha(\beta)$. We extract the scaling dimension $\tau_q$, defined as 
\begin{equation}\label{Eq:tau-def}
  P_q = \sum_\beta \langle |\psi(\beta)|^{2q}\rangle \propto \frac{1}{{\cal D}^{\tau_q}}, 
\end{equation}
where ${\cal D}$ is the Hilbert space dimension, and the brackets denote averaging over disorder and eigenstates. We obtain $\tau_q$ using ED data for systems up to $L=16$ spins, since SI data do not provide a complete set of matrix elements. While fractality is often described using the fractal spectrum $f(\alpha)$ related to $\tau_q$ via the Legendre transform,~\cite{MirlinRMP} we find that $\tau_q$ suffers from fewer numerical issues, and concentrate on its studies below.

Fig.~\ref{Fig:IPR} displays $\tau_q$ for matrix elements of $O=\sigma^z_i$. All curves pass through points $\tau_0=-1$ and $\tau_1 = 0$, which are fixed by the dimension of Hilbert space and the normalization of the wave function. If the distribution of matrix elements is very narrow, all $ |\psi(\beta)|^{2} \propto 1/{\cal D}$, and one expects $\tau_q = q-1$, as illustrated by the diagonal dashed line in Fig.~\ref{Fig:IPR}. At weak disorder $W=0.5$, $\tau_q$ is indeed close to the asymptotic expected for a metal. 

Upon increasing disorder, the spectrum $\tau_q$ stays close to $q-1$ for $q<1$ but begins to deviate for $q>1$. The apparent  in  Fig.~\ref{Fig:IPR} saturation of $\tau_q$ to a constant for $q>1$ corresponds to the termination of fractal spectrum of $f(\alpha)$.~\cite{MirlinRMP} Physically, this is equivalent to a finite probability to find the amplitude of the wave function arbitrarily close to one, that behaves like $1/{\cal D}^c$ with exponent $c$ given by the saturation value of  $\tau_{q}$.  For disorder values in the delocalized phase, this probability vanishes in the thermodynamic limit, corresponding to rare events. Intuitively, such rare events giving very large matrix elements can occur when our local perturbation $\hat O$ acts within a ``localized patch''. 

Finally, roughly when the system enters the MBL phase, all $\tau_q=0$ for $q>q_f$, with $q_f<1$.  Such a spectrum of $\tau_q$ corresponds to the so-called ``frozen'' phase,~\cite{MirlinRMP,Chamon96,Foster14} which combines properties of localized and critical states. In this case there is a \emph{finite} probability of matrix elements being close to one, so there exist a finite number of large matrix elements which saturate the sum $\sum |O_{\alpha\beta}|^2\approx 1$. The remaining matrix elements, which contribute exponentially small weight to the normalization, still have non-trivial behavior corresponding to multifractal fluctuations and correlations.

The frozen phase is a signature of ``local perturbations having strictly local effect"~\cite{Serbyn13-1}, proposed as a defining characteristic of the MBL phase. Indeed, matrix elements creating only local excitations are of order one with finite probability, thus dominating the norm $\sum |O_{\alpha\beta}|^2$. On the other hand, $\tau_q$ with sufficiently small $q$ probes the behavior of increasingly smaller matrix elements which do have non-trivial structure,~\cite{Serbyn15} despite being exponentially suppressed in the system size. Note, that these numerical findings are in apparent disagreement with Ref.~\onlinecite{Montus-frac}, which suggests the scaling dimensions at the MBLT transition to be  $\tau_q = 2q-1$ for $q\leq 1/2$, similar to the case of Anderson transition in infinite dimension.~\cite{MirlinRMP} 

{\it Summary and discussion.---}We summarize our main results which can be viewed as a manifestation of a broad critical regime preceding the MBLT. This region starts from $W_*\approx 2$ for largest accessible system sizes $L=20$. The fact that this region shrinks with increasing $L$ strongly suggests a direct transition between thermal and MBL phases. The critical regime is characterized by: 

{\it (i)}  Matrix elements depend on the energy difference in a way consistent with the ETH. The function $f^2(\omega)$ behaves as a power law in a broad range of energies, $\Eth<\omega<1$. However,  $f^2(\omega)$ ceases to be smooth for $W\geq W_*$, which is manifested by the disparity between average $\phi_{av}$ and typical power $\phi_{typ}$.

{\it (ii)} The many-body Thouless energy, extracted from the spectral function $f^2(\omega)$, satisfies $\Eth\gg \Delta$  and  obeys the scaling consistent with subdiffusion at weak disorder. However, for $W\geq W_*$ the Thouless energy remains comparable to the level spacing for the largest system sizes.

{\it (iii)} Interpreting the matrix elements as the wave function of a local excitation, we find multifractal behavior. For $W>W_*$ the scaling dimensions of the participation ratios $\tau_q$ begin to deviate from the value in a metal, both at $q>1$ and $q<1$. MBL transition roughly agrees with the onset of the ``frozen fractal spectrum''.

{\it Acknowledgments.---}We acknowledge useful discussions with V. Kravtsov,  T. Grover, and R. Vasseur. 
M.S. was supported by Gordon and Betty Moore Foundation's EPiQS Initiative through Grant GBMF4307. M.S. and D.A. acknowledge hospitality of KITP, where parts of this work were completed (supported in part by the National Science Foundation under Grant No. NSF PHY11-25915).

%

\pagebreak
\clearpage
\appendix 
\section{Energy dependence of matrix elements}

\begin{figure}[b]
\includegraphics[width=.99\columnwidth]{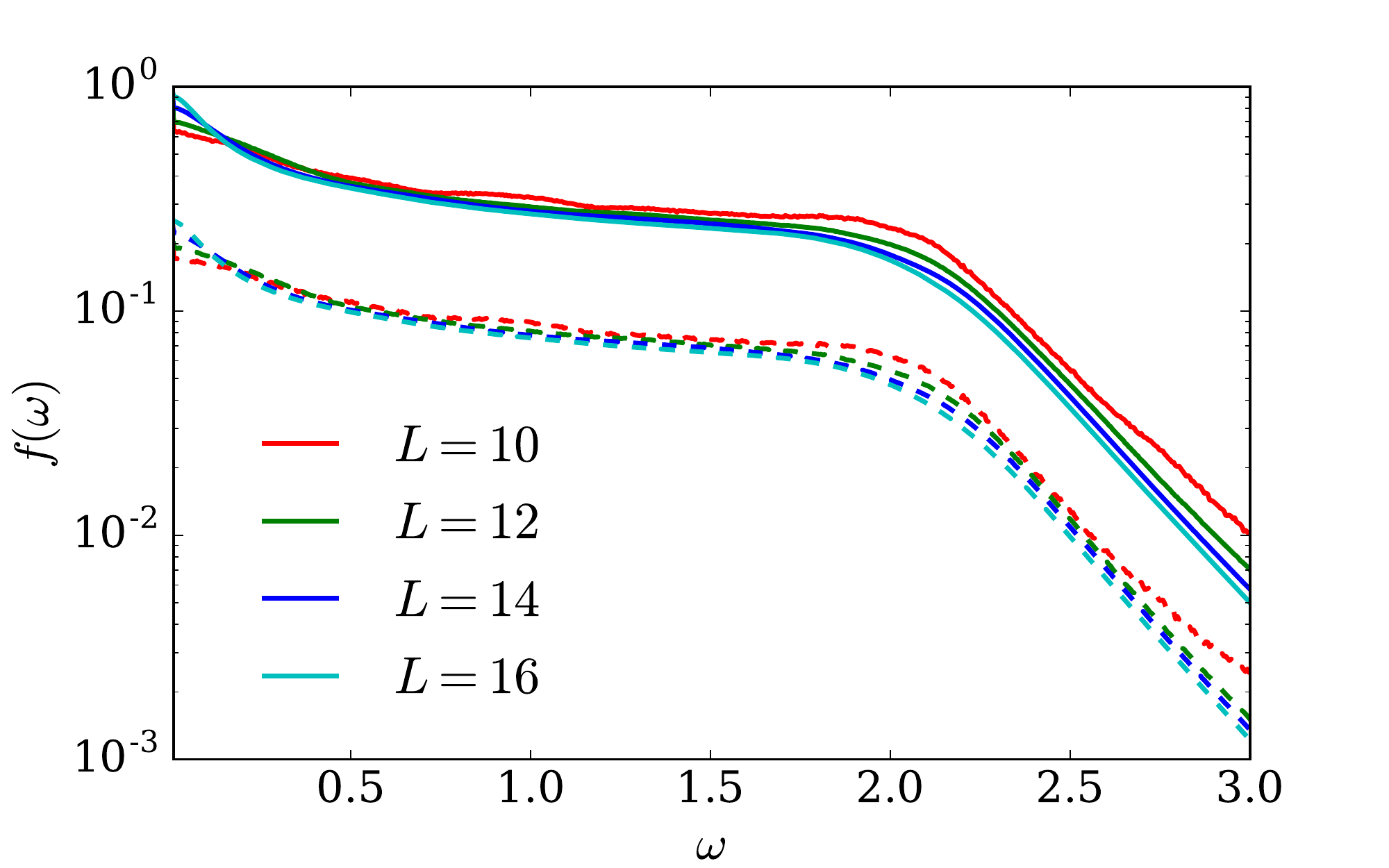}
\includegraphics[width=.99\columnwidth]{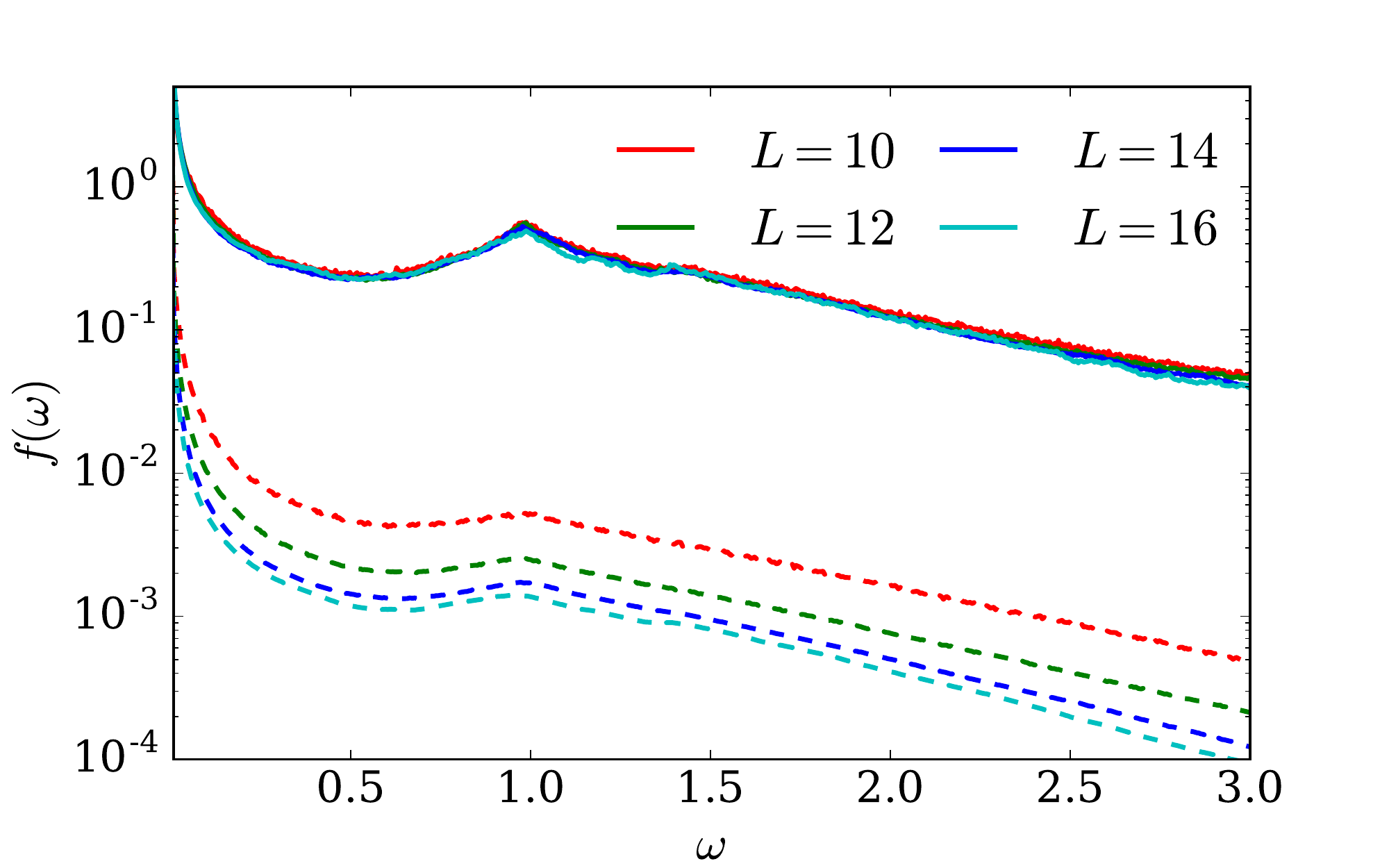}\\
\caption{ \label{Fig:f-global} (Color online) Top: For weak disorder ($W=0.5$), $f^2(\omega)$ for different system sizes collapse onto each other for both average (solid lines) and typical (dashed) curves. 
Bottom: For disorder $W=2.75$, the collapse breaks down for typical curves. The Thouless energy, that was visible in the top plot at small $\omega$, now cannot be resolved, and solid lines collapse onto each other in the full range.}
\end{figure}

Below we present additional numerical data to support the conclusions in the main text. We start by showing $f^2(\omega)$ as a function of $\omega$ in Fig.~\ref{Fig:f-global}. This allows us to clearly illustrate the exponential decay of $f^2(\omega)$ when $\omega$ is of order of microscopic energy scale. In addition, in the top panel of Fig.~\ref{Fig:f-global} one can see that $f^2(\omega)$ do not collapse onto each other for different system size near $\omega=0$, which is the manifestation of the Thouless energy  decreasing with system size. In contrast, at disorder closer to the MBLT, the Thouless energy becomes smaller than the level spacing, and hence the curves in the bottom panel of Fig.~\ref{Fig:f-global} perfectly collapse onto each other down to the smallest value of $\omega$. 

\begin{figure}[b]
\includegraphics[width=.99\columnwidth]{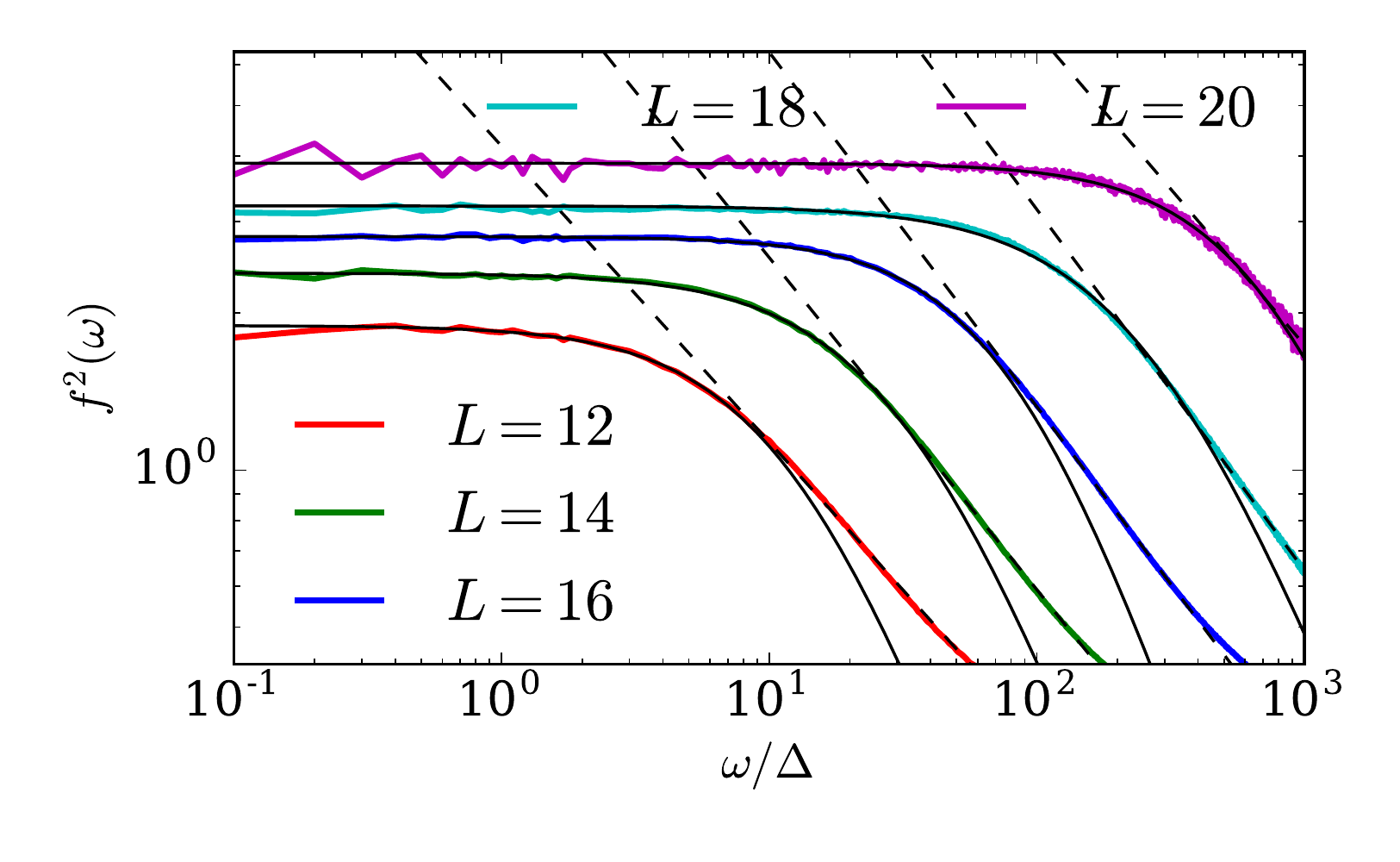}
\includegraphics[width=.99\columnwidth]{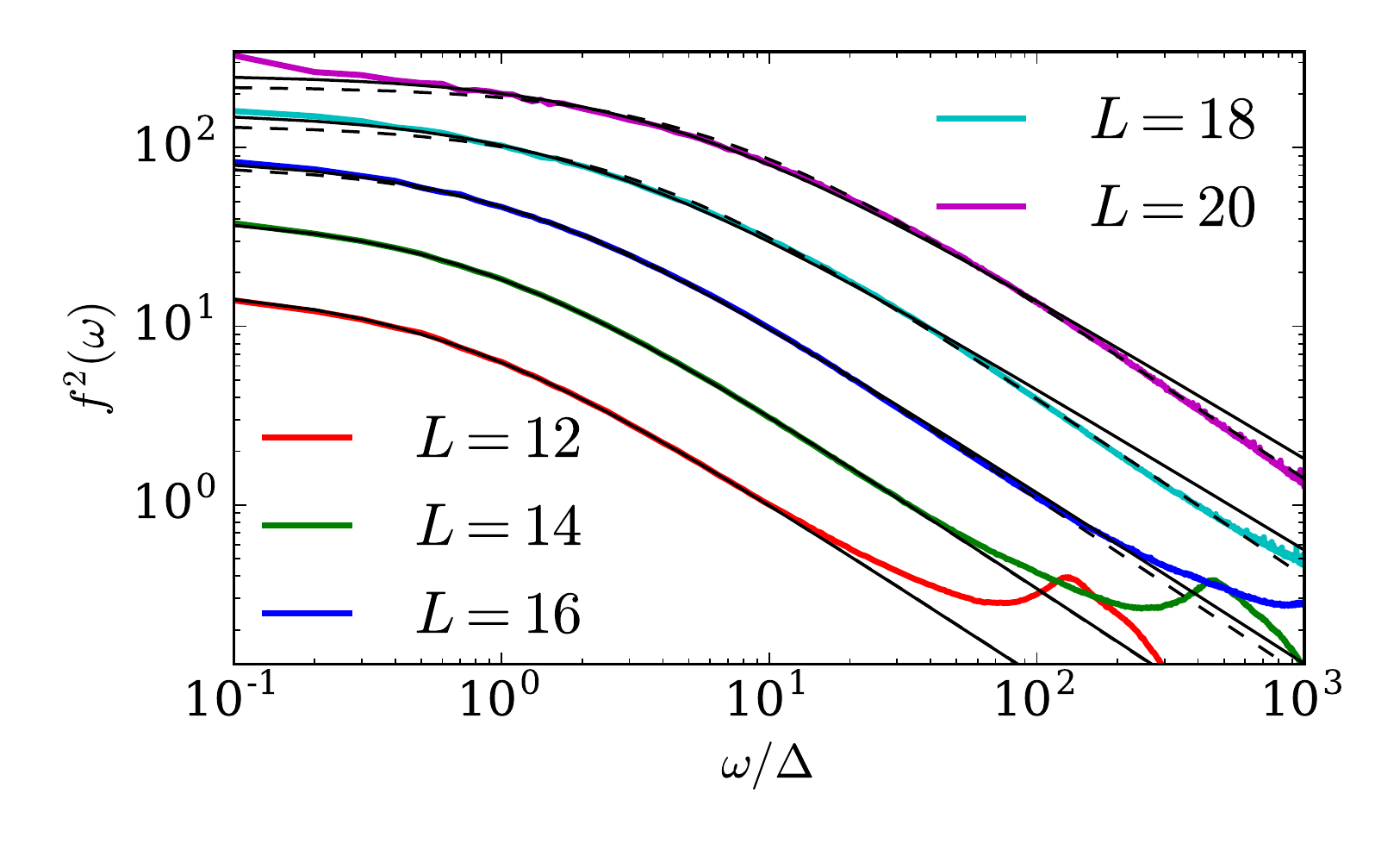}\\
\caption{ \label{Fig:fits} (Color online) Illustration of the fits quality of $f^2(\omega)$ for $W=1$ (top panel) and $W=2$ (bottom panel). Note, that Eq.~(\ref{Eq:fit}) fails to capture the onset of decay of $f^2(\omega)$ at weak disorder in the top panel, thus we fitted the data with the power-law (straight dashed lines). Dashed lines denote fits used to extract the power $\phi$, whereas solid lines were used to determine $\Eth$ and $f^2(0)$.}
\end{figure}
In order to demonstrate our fitting procedure, we show the fits used to extract the power $\phi$, as well as the Thouless energy in Fig.~\ref{Fig:fits}. We see that at $W=1$, Eq.~(\ref{Eq:fit}) fails to adequately capture the decay of $f^2(\omega)$; therefore for disorders $W=1$ and $W=1.5$ we extracted $\phi$ by fitting the ``shoulder'' to the power-law behavior, as is illustrated by dashed lines in the top panel of Fig.~\ref{Fig:fits}. However, already for disorder $W\geq 1.75$ Eq.~(\ref{Eq:fit}) adequately describes the behavior of $f^2(\omega)$. In particular, bottom panel of Fig.~\ref{Fig:fits} shows excellent agreement between $f^2(\omega)$ and its fits by Eq.~(\ref{Eq:fit}). For all values of the disorder $W\geq 1.75$  we used fits of $f^2(\omega)$ to extract the Thouless energy and $f^2(0)$. In order to capture the power-law decay more accurately, we fitted $\ln f^2(\omega)$ to the logarithm of Eq.~(\ref{Eq:fit}); corresponding fits are shown by dashed lines in the bottom panel of Fig.~\ref{Fig:fits}.

\begin{figure}[htb]
        \includegraphics[width=0.99\columnwidth]{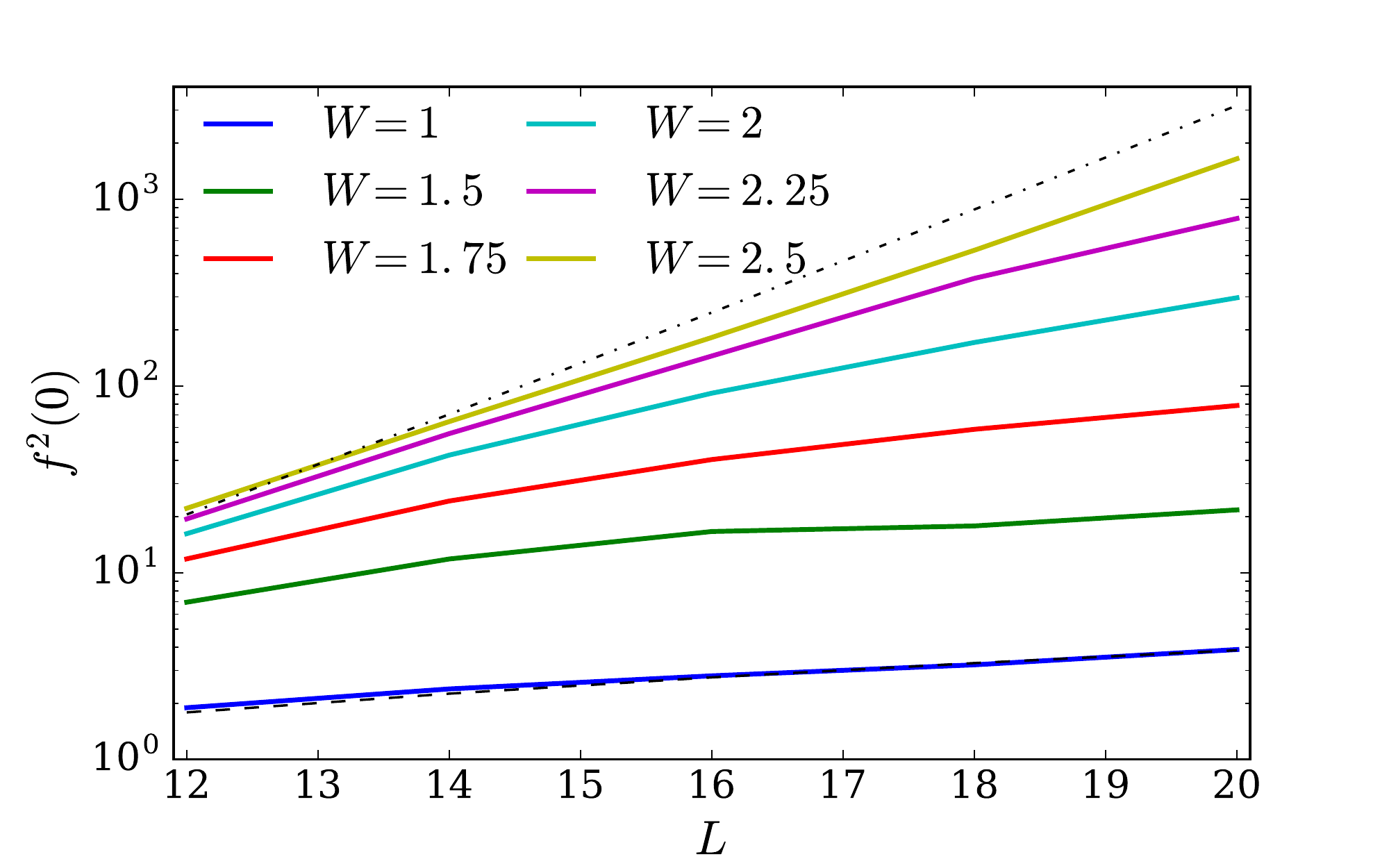}
        \caption{\label{Fig:f2-sum-rule} (Color online)
Dependence of $f^2(0)$ on $L$ for different values of the disorder. For weak disorder, it is consistent with the power-law dependence of $\Eth$ on the system size (the bottom dashed line corresponds to diffusive behavior). Upon increasing the disorder, $f^2(0)$ gets enhanced, but still grows slower that $1/\Delta$ which is illustrated by the top dot-dashed line.  }
\end{figure}
Next, we show the magnitude of $f^2(0)$, extracted from the fits of $f^2(\omega)$ in Fig.~\ref{Fig:f2-sum-rule}, which is consistent with the dependence of $\Eth$ on system size, shown in Fig.~\ref{Fig:Eth} in the main text. Indeed, if one assumes that the expectation value of $\hat O$ vanishes with system size, $f^2(\omega)$ obeys the sum rule, so that the behavior of $f^2(0)$ can be related to the finite size scaling of $\Eth$:~\cite{Polkovnikov-rev}  from $\Eth\propto L^{-1/\gamma}$ we get $f^2(\omega)\propto L^{1/\gamma-1}$, e.g., it increases with system size in a power-law fashion. In contrast, if the Thouless energy is comparable to the level spacing, we expect that $f^2(\omega)$ scales with the inverse level spacing, as is the case in Fig.~\ref{Fig:f2-sum-rule} for larger values of the disorder. 

We note that this suggests an alternative interpretation of the results in recent  Ref.~\onlinecite{Luitz-fluc-16}. Our Fig.~\ref{Fig:f2-sum-rule} suggests that while the power-law scaling of $\mathop{\mathrm{std}} (O_{\alpha\beta} e^{S/2}) $ holds at weak disorder, for stronger disorder within the critical region [i.e., for $W\geq W_*$], the exponential dependence $\mathop{\mathrm{std}} (O_{\alpha\beta} e^{S/2}) \propto e^{\alpha L}$ is more natural. This may explain a strong upward curvature of  $\mathop{\mathrm{std}} (O_{\alpha\beta} e^{S/2}) $  plotted as a function of $L$ on the double logarithmic scale in Fig.3 of~Ref.~\onlinecite{Luitz-fluc-16}.

\begin{figure}[b]
\includegraphics[width=0.99\columnwidth]{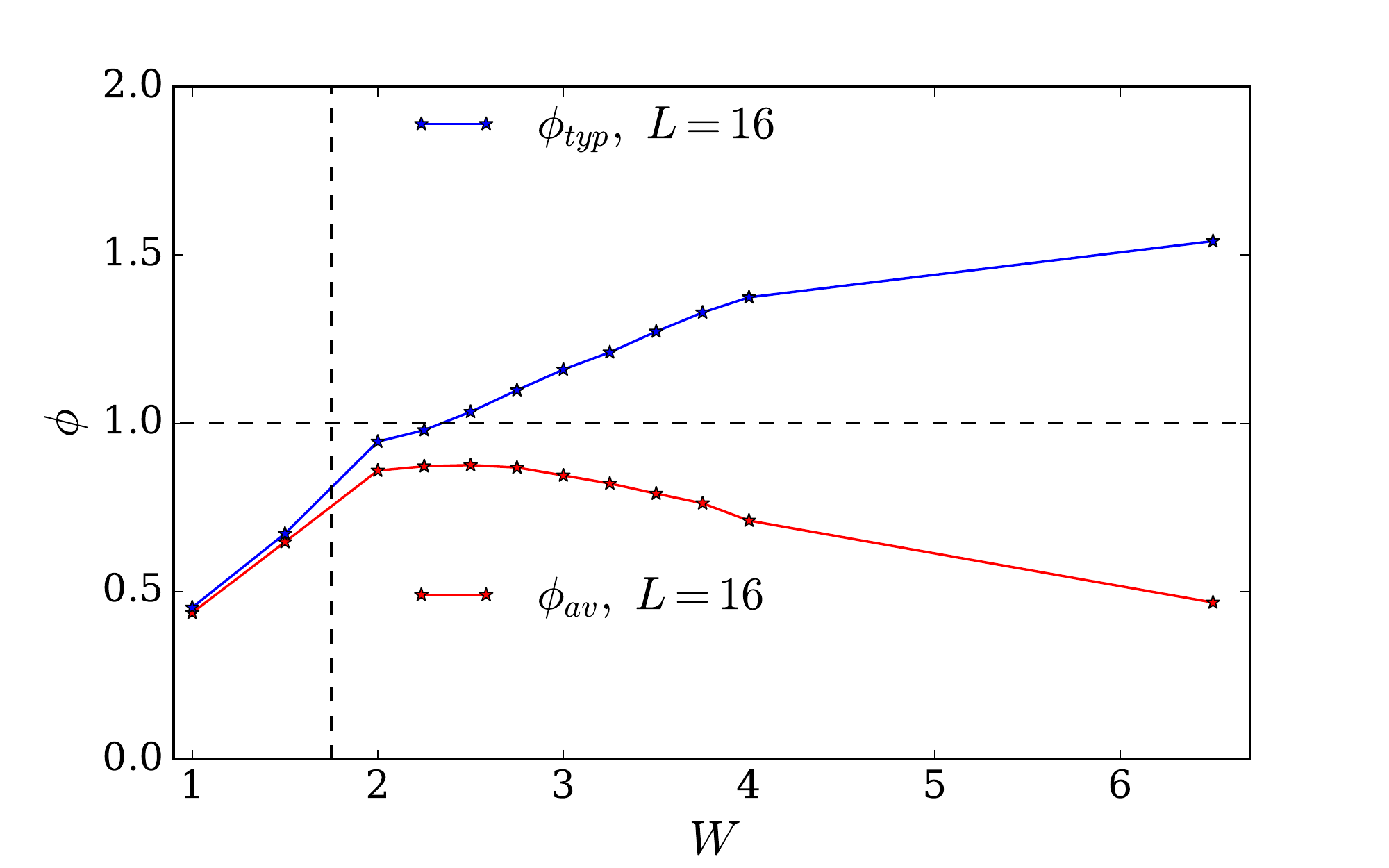}
\caption{ \label{Fig:SpSm} (Color online) Average and typical exponents $\phi$ governing the decay of $f^2(\omega)$ behave in a qualitatively similar way for the operator $\sigma^x_i\sigma^x_{i+1}+\sigma^y_i\sigma^y_{i+1}$.}
\end{figure}
Finally, Fig.~\ref{Fig:SpSm} provides the data for the exponent governing the power-law decay for the matrix elements of the operator $\sigma^x_i\sigma^x_{i+1}+\sigma^y_i\sigma^y_{i+1}$. Similar to the matrix elements of $\sigma^z$ operator, shown in Fig.~\ref{Fig:phi}, the power $\phi$ in Fig.~\ref{Fig:SpSm} freezes at a value close to one when $W\geq W_1$. At the same time, the power governing the decay of log-averaged spectral function, $\phi_{typ}$ begins to deviate from~$\phi$. 

\section{Fractal properties of matrix elements}

\begin{figure}[t]
       \includegraphics[width=.99\columnwidth]{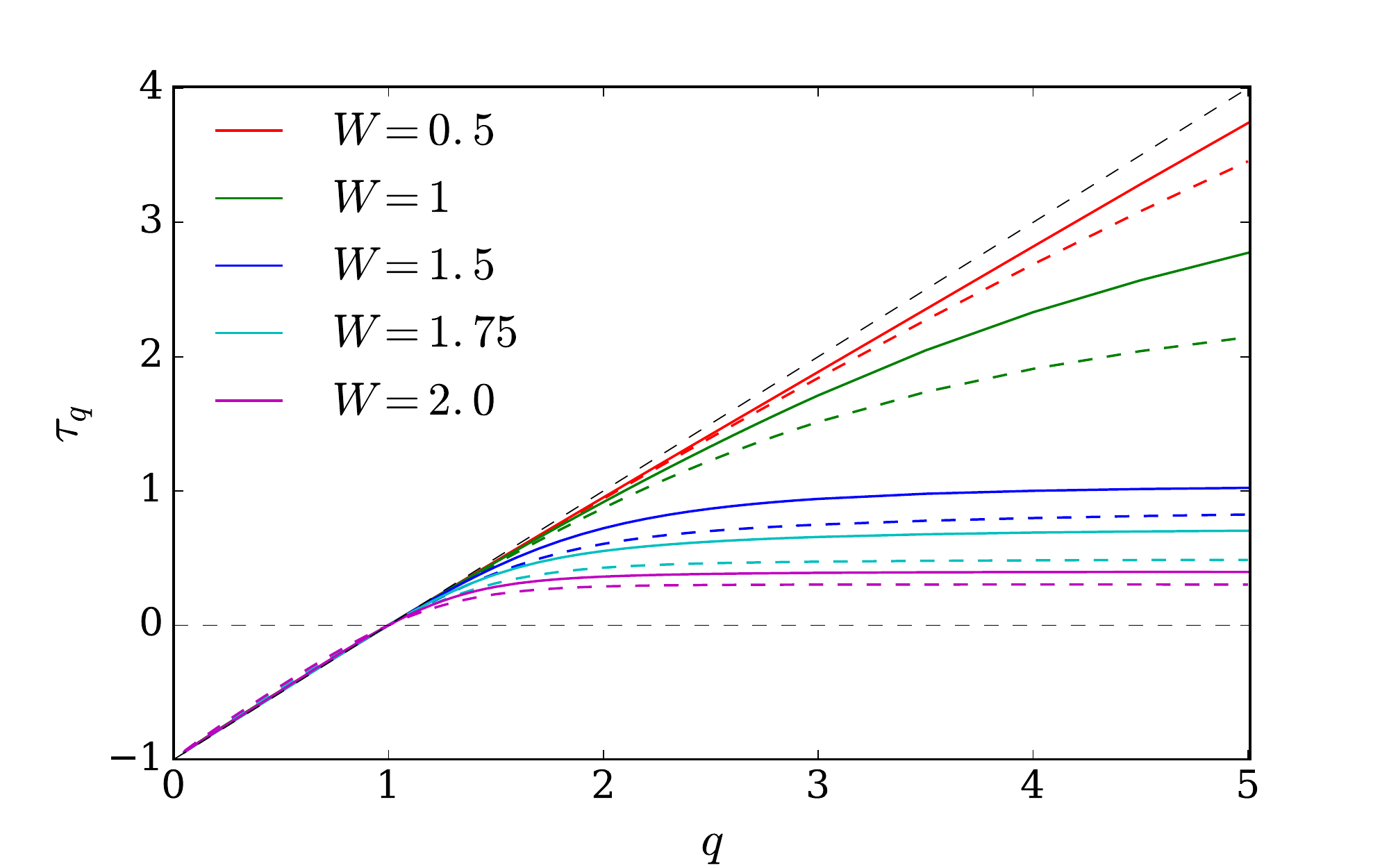}
\caption{ \label{Fig:IPR-app} (Color online) The scaling dimension $\tau_q$ for $\hat O   = \sigma^z_i$, obtained by fitting $P_q$ from fits for $L=10\ldots 16$ (solid lines) and $L=8\ldots 14$ (dashed lines). Note that while $\tau_q$ is always closer to its metallic value for larger system sizes, the flow becomes much slower for $W\geq 2$.}
\end{figure}

Our data for the scaling dimension, Fig.~\ref{Fig:IPR} in the main text, illustrated the presence of fractality even at very moderate disorders. Here we address the finite size scaling of these scaling dimension: Fig.~\ref{Fig:IPR-app} shows $\tau_q$ extracted from smaller (dashed lines) and larger (solid lines) system sizes. 

While $\tau_q$ consistently deviates below $q-1$ for $q>1$ even at disorder $W=0.5$, we see a strong flow of $\tau_q$ with system size, and $\tau_q$ becomes closer to $q-1$ for larger system sizes. In contrast, when the disorder approaches $W_*$, so that we are in the critical region near the MBLT, the finite size flow of $\tau_q$ becomes hardly noticeable. This supports the conclusion that the fractal behavior of the matrix elements is pertinent to the wide critical region surrounding the MBLT. 

\begin{figure}[b]
      \includegraphics[width=.99\columnwidth]{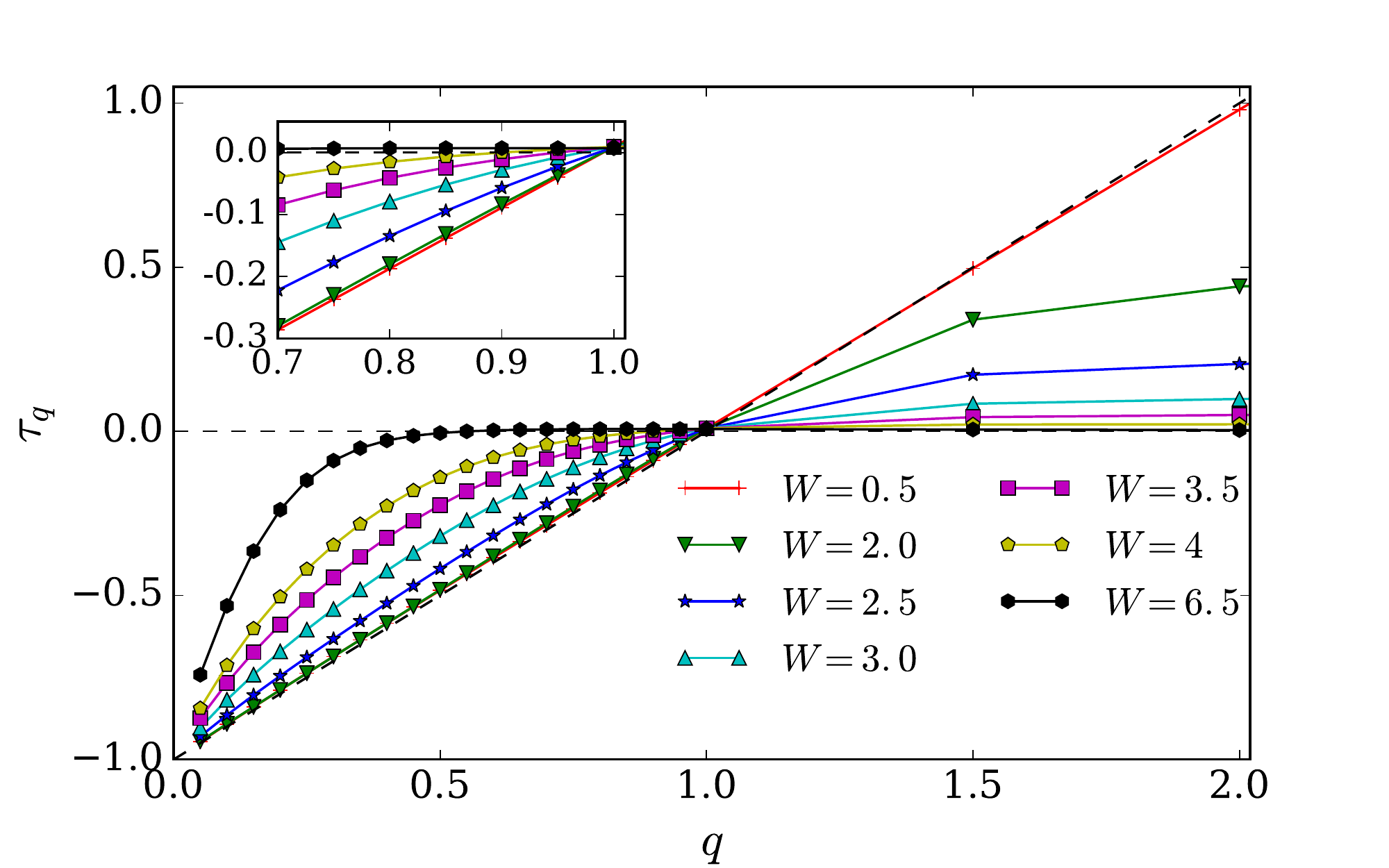}
\caption{ \label{Fig:IPR-app2} (Color online) Scaling dimension for the operator $\hat O = \sigma^x_i\sigma^x_{i+1}+\sigma^y_i\sigma^y_{i+1}$, flipping spins at adjacent sites when they point in  different directions.}
\end{figure}
\begin{figure*}[t!]
\includegraphics[width=.66\columnwidth]{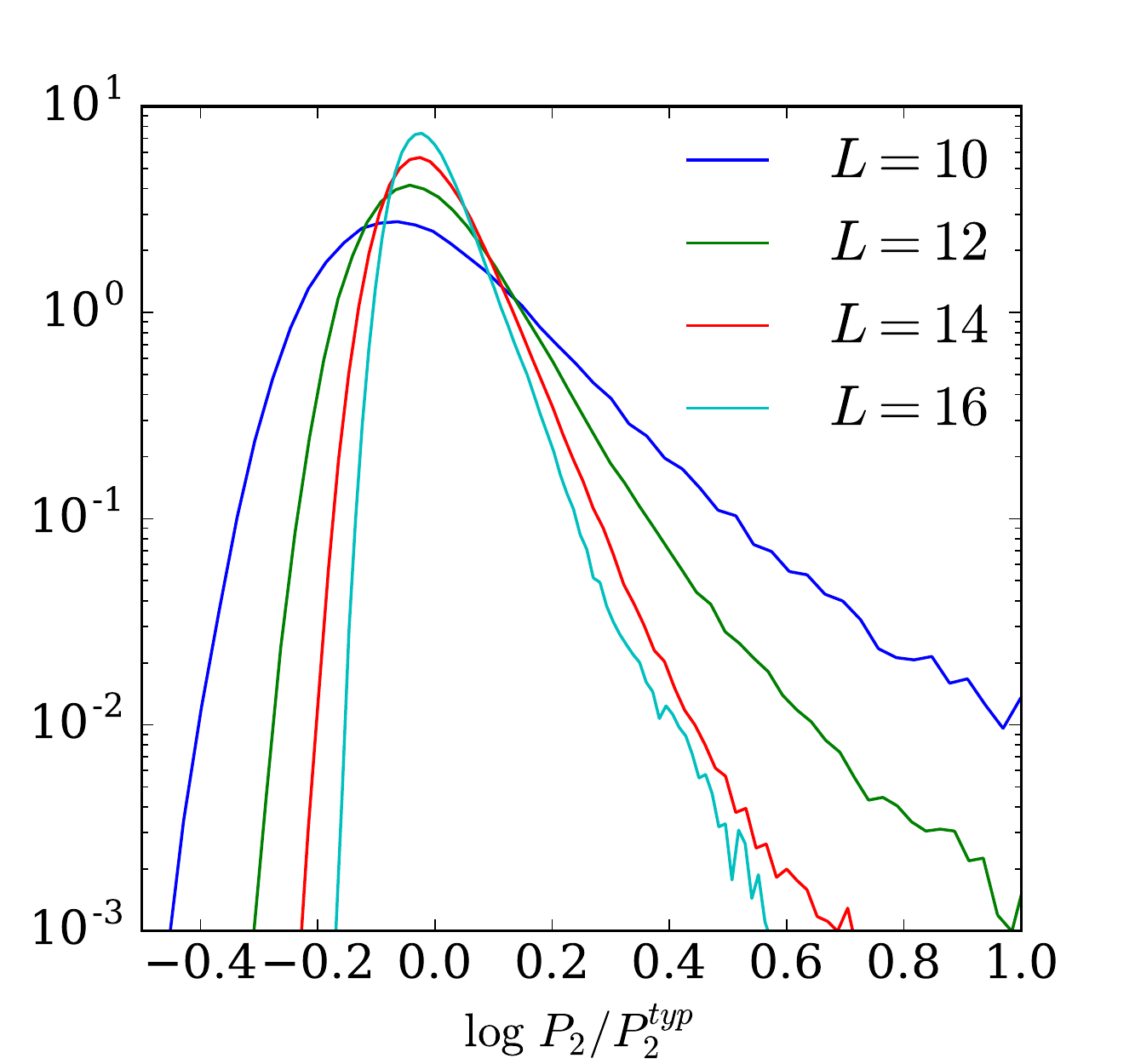}
\includegraphics[width=.66\columnwidth]{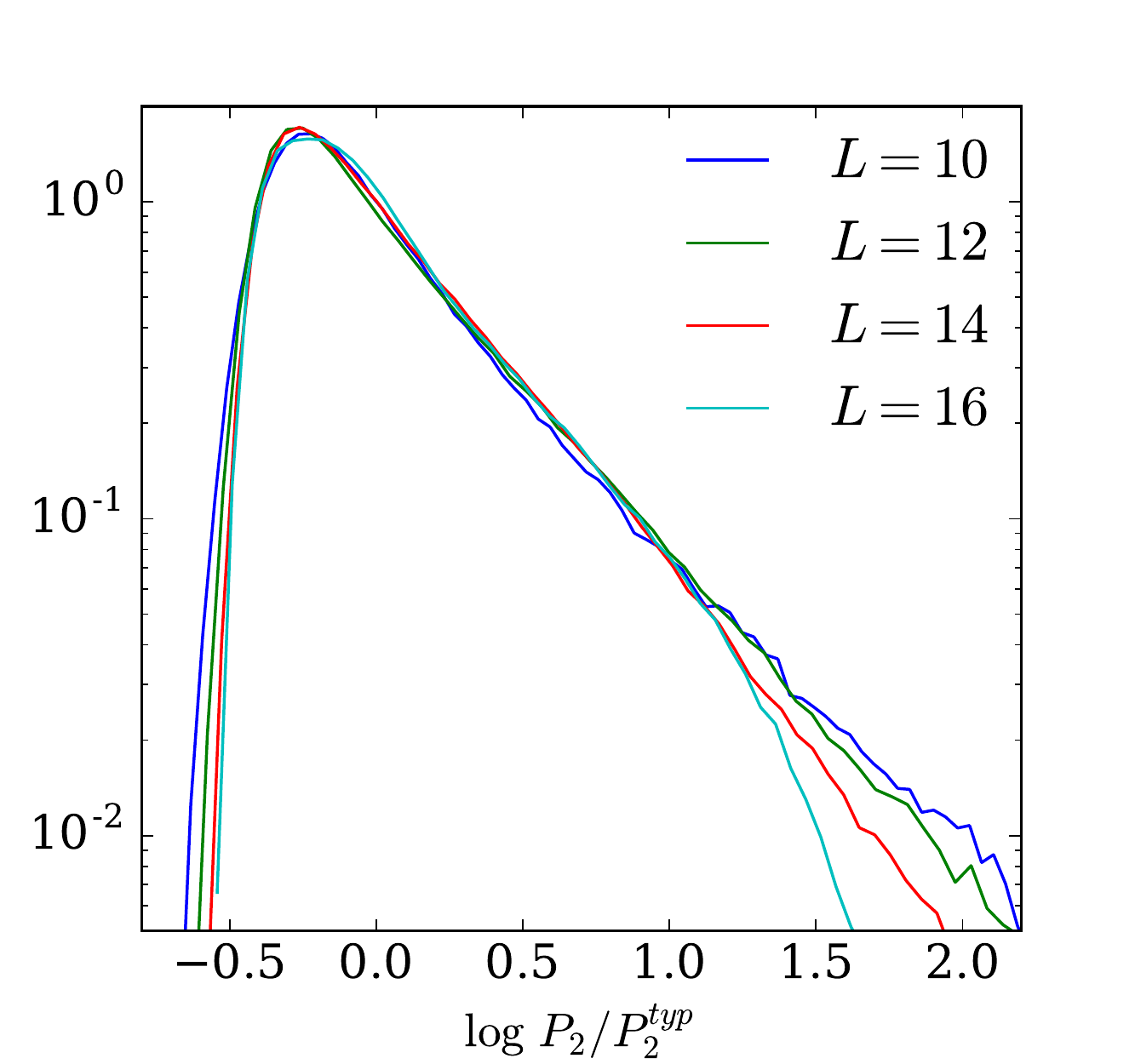}
\includegraphics[width=.66\columnwidth]{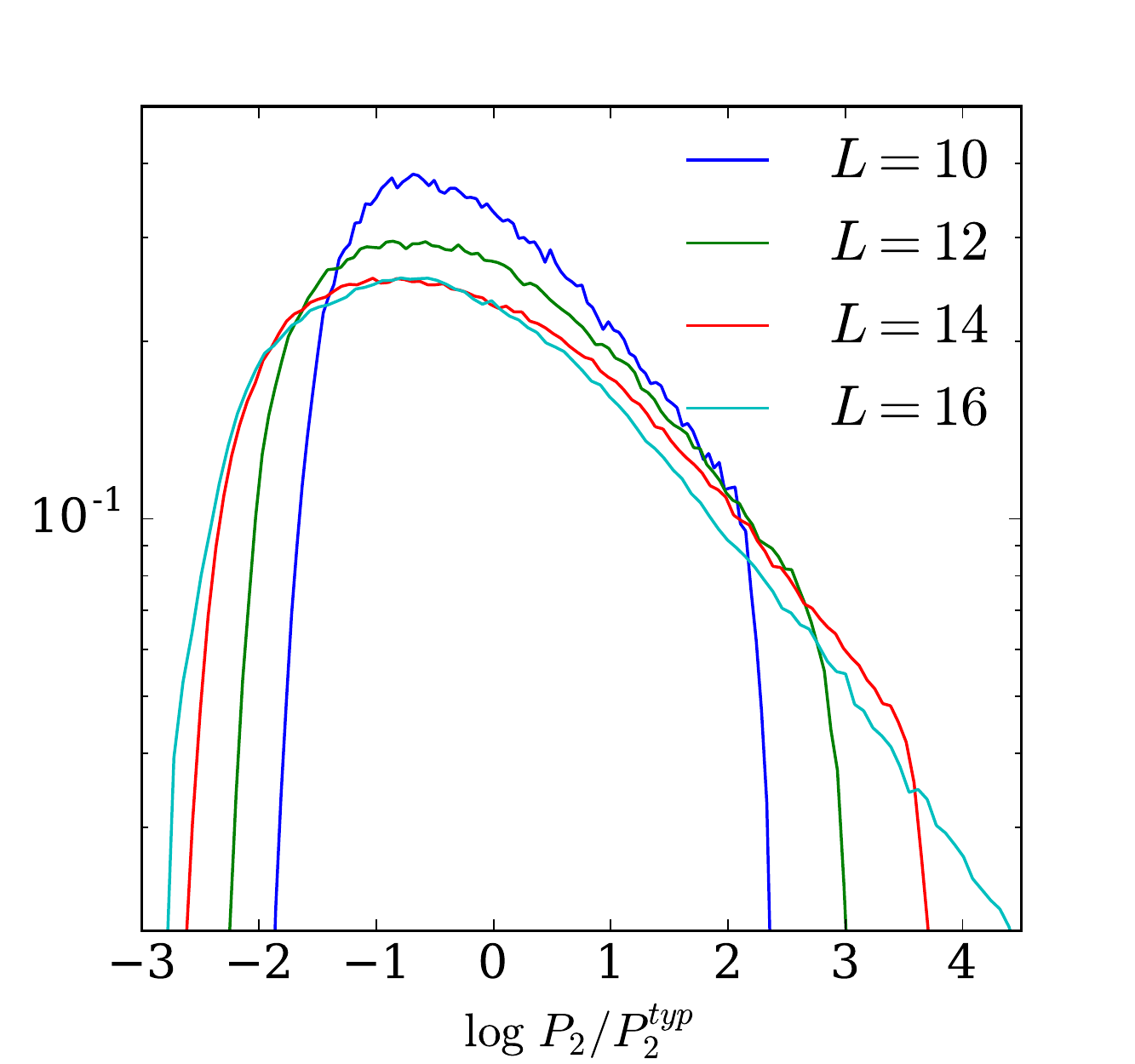}
\caption{ \label{Fig:IPR-dist} (Color online) Distribution of participation ratios $P_2$ on the ergodic side of the MBLT. For weak disorder, $W=0.5$ (left), the distribution becomes increasingly narrower for larger systems. However, already for disorder $W=1$ (middle) the distribution looks scale-invariant, with the power law decay of probability to have $P_2>P_2^{typ}$. Finally, at even stronger disorder $W=2$ (right), the distribution consistently broadens with system size. }
\end{figure*}

Next, we address the fractality for the case of a different operator -- namely we consider the operator $\sigma^x_i\sigma^x_{i+1}+\sigma^y_i\sigma^y_{i+1}$, which flips the spins at two adjacent sites when they point in opposite directions. While such an operator no longer squares to the identity, we see that the condition $\tau_1=0$ still holds, suggesting that the squared sum of the matrix elements $\sum_\beta O_{\alpha\beta}^2$, does not change significantly with the system size.  Scaling dimensions for this spin-flip operator show the behavior similar to the case of $\sigma^z$ operator: $\tau_q$ weakly deviates from the metallic behavior at weak disorder, then consistent with the termination of the fractal spectrum, $\tau_q$ saturates for some $q>1$. Finally, in the MBL phase we also observe the characteristic ``frozen'' spectrum.

 Finally, we also present the distribution of the participation ratio of matrix elements of the operator $\sigma^z$ in Fig.~\ref{Fig:IPR-dist}. At weak disorder, the width of the distribution shrinks, consistent with $\tau_q$ approaching the limit $q-1$ of the ideal metal. However, already at disorder $W=1$ we observe the approximately scale invariant form of the above distribution, with the power-law decay of probability to have large participation ratios $p(P_2/P_2^{typ})\propto 1/P^{1+x_2}_2$. For $W=1$ we have $x_2\approx 2$. Exponent $x_2$ decreases for increasing disorder, and it is close to 1 already for $W=1.5$. Note, that when $x_2<1$ the scaling dimension extracted from $P_2$ begins to deviate from the ``typical'' scaling dimension extracted from log-averaged participation ratio, see Ref.~\onlinecite{MirlinRMP} for more details. Finally, at even stronger disorder, $W=2$, the distribution of participation ratios loses its scale invariant form and broadens for larger systems.

\end{document}